\documentclass{article}

\usepackage{arxiv}
\usepackage[utf8]{inputenc} 
\usepackage[T1]{fontenc}    
\usepackage[hidelinks]{hyperref}       
\usepackage{url}            
\usepackage{booktabs}       
\usepackage{amsfonts}       
\usepackage{nicefrac}       
\usepackage{microtype}      
\usepackage{lipsum}		
\usepackage{graphicx}
\usepackage{doi}
\usepackage{cleveref}
\usepackage[
    backend=biber,
    style=numeric-comp,
    ]{biblatex}
\usepackage{wrapfig}
\usepackage[labelfont=bf]{caption}
\usepackage{subcaption}
\usepackage{tocloft}

\title{Letter of Interest: Ocean science with the Pacific Ocean Neutrino Experiment}
\newcommand{\shorttitle}{Letter of Interest: P-ONE Ocean Science}
\author{%
    F. Henningsen \\
    Department of Physics \\
    Simon Fraser University \\
    Burnaby, BC, V5A 1S6, Canada \\
    \href{mailto:felix_henningsen@sfu.ca}{\tt felix\_henningsen@sfu.ca} \\
    \And
    L. Schumacher \\
    Department of Physics \\
    Technical University Munich \\
    Garching, 85748, Germany \\
    \href{mailto:lj.schumacher@tum.de}{\tt lj.schumacher@tum.de} \\
	\AND
	for the P-ONE Collaboration
}

\date{}

\begin{document}
\maketitle

\begin{abstract}
    The Pacific Ocean Neutrino Experiment (P-ONE) is a proposed cubic-kilometer scale neutrino telescope planned to be installed in the deep-sea of the north-east Pacific Ocean. In collaboration with the optical deep-sea data and communications network operated by Ocean Networks Canada, an international collaboration of researchers plans to install an array of kilometer-long mooring lines in a depth of around $2660\,$m to the relatively flat deep-sea region called Cascadia Basin, around 300 miles West of Vancouver Island. With the design and development ongoing, the P-ONE collaboration is interested to initiate participation of fellow scientists of the oceanographic and marine science communities to provide expertise and experience towards deploying additional or inclusive instrumentation and measurement strategies for doing oceanographic research. In addition to the monitoring of optical bioluminescence and deep-ocean dynamics and thermodynamics, active and passive acoustics can be installed within the P-ONE array. This letter summarizes the P-ONE detector and a non-exhaustive list of potential topics of interest for oceanographic and marine research.
\end{abstract}

\vspace*{1cm}

\hrule

\setlength{\cftbeforetoctitleskip}{-1em}
\renewcommand\contentsname{}
\vspace*{1cm}
\tableofcontents
\vspace*{1cm}

\section{P-ONE Interdisciplinary Meeting}
On \textbf{September 30}, the second P-ONE Interdisciplinary Meeting will take place in \textbf{Vancouver, Canada}, hosting the already-involved research community of oceanographers and marine scientists. If you are curious to participate and see what options there are to get involved with the experiment, feel free to reach out to the authors for information on how to participate. Virtual participation can be offered to anyone interested, in-person attendance is possible, however, travel reimbursements by the collaboration are not available at this time.

\begin{center}
    \large
    \textbf{P-ONE Interdisciplinary Meeting}\\[4pt]
    \textbf{Date} September 30, 2022\\[4pt]
    \textbf{Location} Vancouver, BC, Canada\\[4pt]
\end{center}

\paragraph*{Objectives}
The Interdisciplinary Meeting will outline the concrete steps for the involvement of external research groups in complementary science initiatives intending to utilize the future P-ONE system. The meeting will therefore be an opportunity for interested researchers to define requirements in terms of sensors \& instruments and data products. The requirements will lead to feasibility and affordability discussion, and to the definition of possible constraints on the system design, including but not limited to, hardware and software interfaces. 

\section{The Pacific Ocean Neutrino Experiment}
\begin{figure}[h!]
    \centering
    \includegraphics[width=0.8\textwidth]{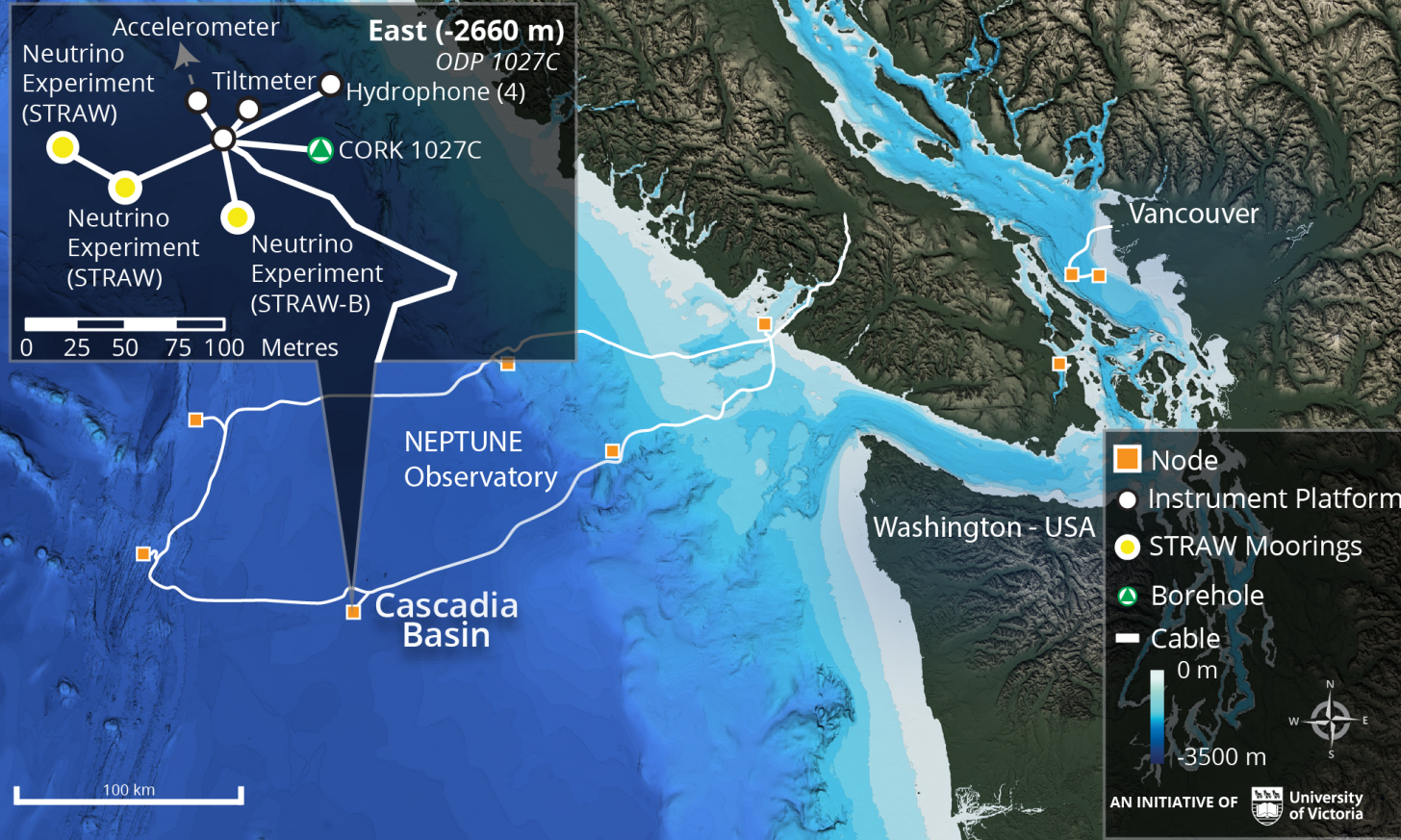}
    \caption{The optical-electrical deep-sea cable network operated by Ocean Networks Canada off the coast of Vancouver Island with its various main connection nodes (orange). The P-ONE detector will be located near the Cascadia Basin node towards the South of the network. Image provided by Ocean Networks Canada. Figure taken from~\cite{Bailly_2021}.}
    \label{fig:map}
\end{figure}

The \textbf{Pacific Ocean Neutrino Experiment}~\cite{Agostini_2020}, short P-ONE, will be a large-scale optical detector for astrophysical neutrinos located on the Juan de Fuca Plate in the north-east Pacific Ocean. The site, called the Cascadia Basin, is located off the West coast of Vancouver Island and is instrumented with a large deep-sea network of optical-electrical cables operated and maintained by \textbf{Ocean Networks Canada} (ONC)\footnote{\href{https://www.oceannetworks.ca/}{https://www.oceannetworks.ca/}}. This research-driven network already provides the means for numerous marine and oceanographic research studies, and will be the infrastructure provider for the P-ONE detector as shown in \cref{fig:map}.

\subsection{Pathfinder missions}
In order to verify that the optical properties of the Cascadia Basin site are sufficient for the physics targets of P-ONE, two pathfinder missions have been deployed to the site: STRAW~\cite{Boehmer_2019} in 2018 and STRAW-b~\cite{Rea_2021o3} in 2020 (STRings for Absorption length in Water). As of 2022, both pathfinder experiments are still operating and sending data to shore.

STRAW and STRAW-b consist of photosensitive detectors attached to mooring lines of $150\,$m and $450\,$m length, respectively. Both of them aim to characterize the P-ONE site in terms of optical properties and light background as a function of time. It was these experiments that proved the water of Cascadia Basin to be optically clear enough to host a large-volume neutrino telescope. Furthermore, the long-term monitoring of bioluminescence activity became a major interest during the operation and oceanographic and marine science researchers are involved in analyzing these data, all in collaboration with ONC~\cite{Bailly_2021}.

\subsection{Detector design}
\begin{figure}[h!]
    \centering
    \includegraphics[width=0.8\textwidth]{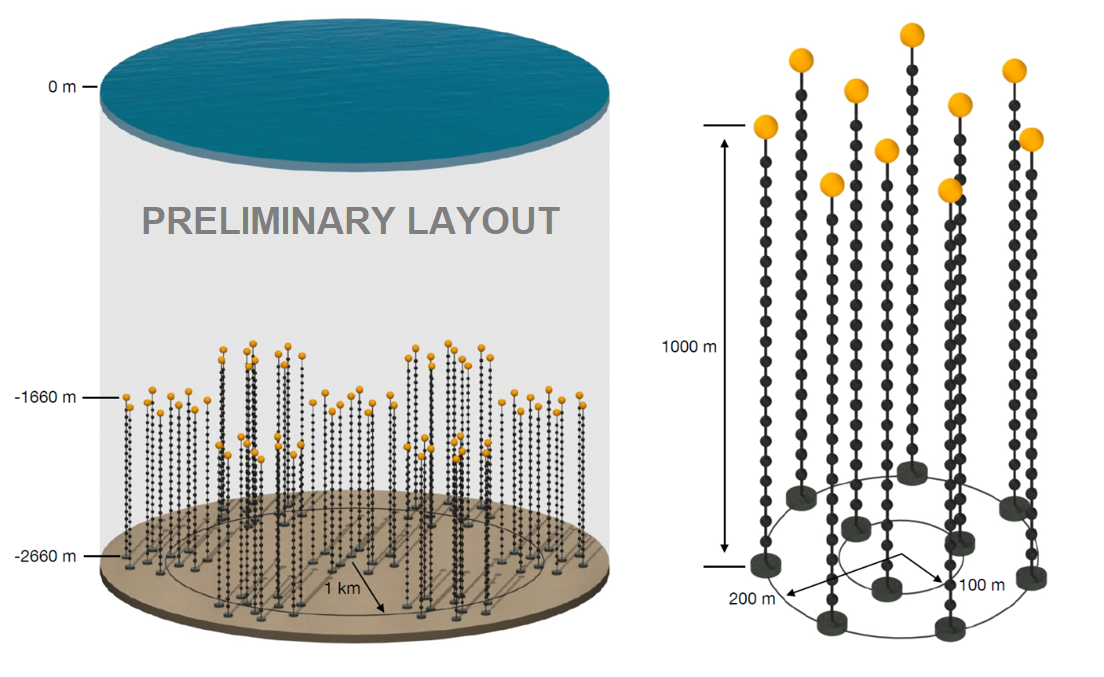}
    \caption{Preliminary P-ONE detector layout for the full array (left) and one cluster of 10 mooring lines (right). The detector spans a height of one kilometer, starting on the sea floor at around $2660\,$m depth. The final detector layout is currently under study. Figure adapted from~\cite{Agostini_2020}.}
    \label{fig:pone}
\end{figure}

The P-ONE detector will be composed of several \textbf{moored observatories} (MOs) arranged in clusters. One cluster will comprise 10 individual MOs, each an instrumented mooring line of one kilometer length, starting on the sea floor at $2660\,$m depth. Each MO will host 20 instruments for collecting light or for calibration purposes, an anchor at the bottom and a buoy on top to keep the MO upright. The instruments itself are enclosed in glass spheres and are individually connected via the \textbf{load-bearing electric-optical backbone cable} to a \textbf{mini junction box} (MJB) which provides power and establishes communication to shore. One MJB is integrated at the bottom of each mooring line, connecting all instruments to the network. Each MO will be mounted to a tray, acting both as the anchor and deployment base for the string.
Active acoustic transmitters for position calibration of the MO will be deployed onto the nearby seafloor. 
An illustration of the project layout is depicted in~\cref{fig:pone}, but note that the final layout is still under investigation.
%
\begin{figure}[h!]
    \centering
    \begin{subfigure}[b]{0.25\textwidth}
        \centering
        \includegraphics[width=1\textwidth]{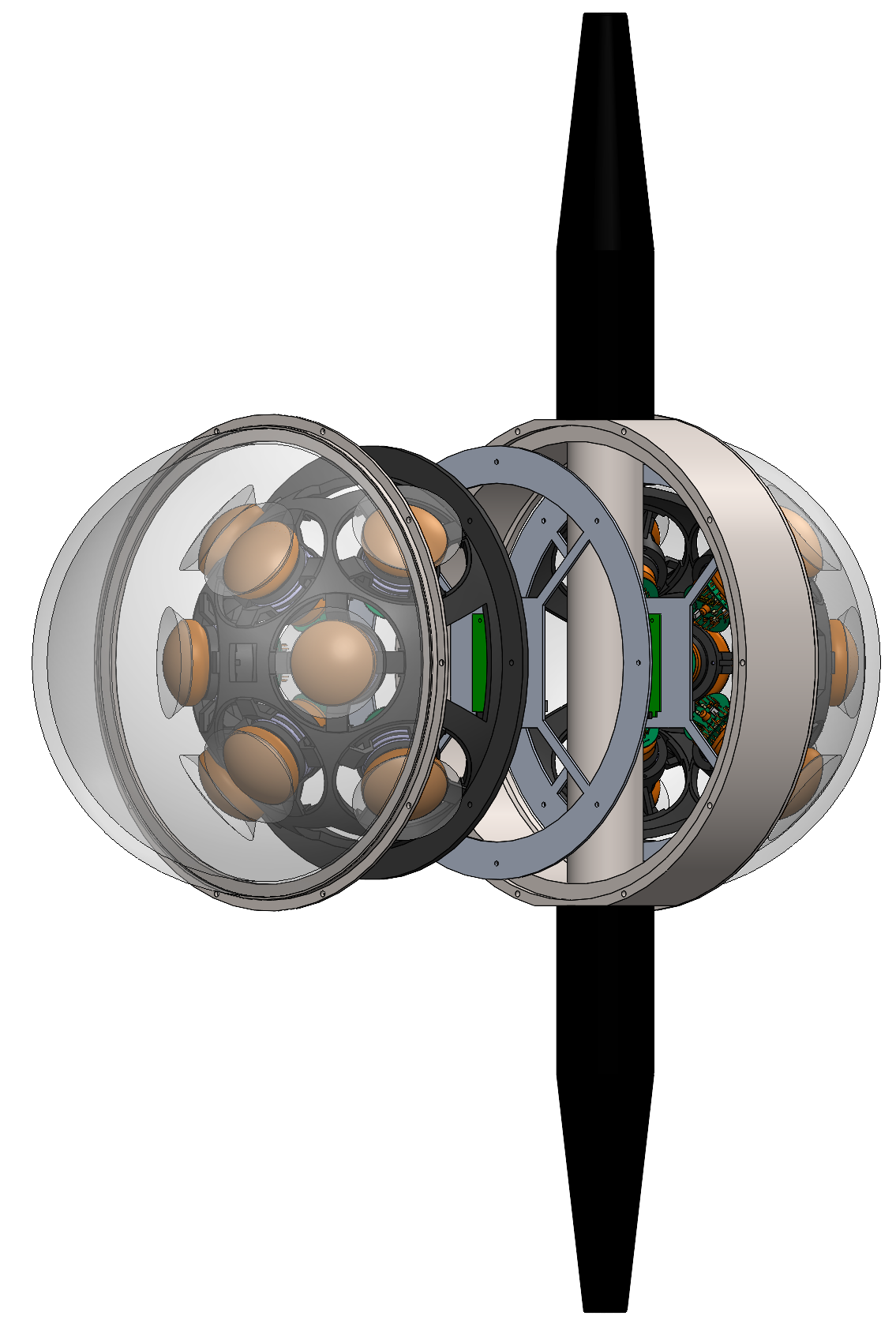}
        \caption{P-ONE cable segment.}
    \end{subfigure}
    \hfill
    \begin{subfigure}[b]{0.74\textwidth}
        \centering
        \includegraphics[width=1\textwidth]{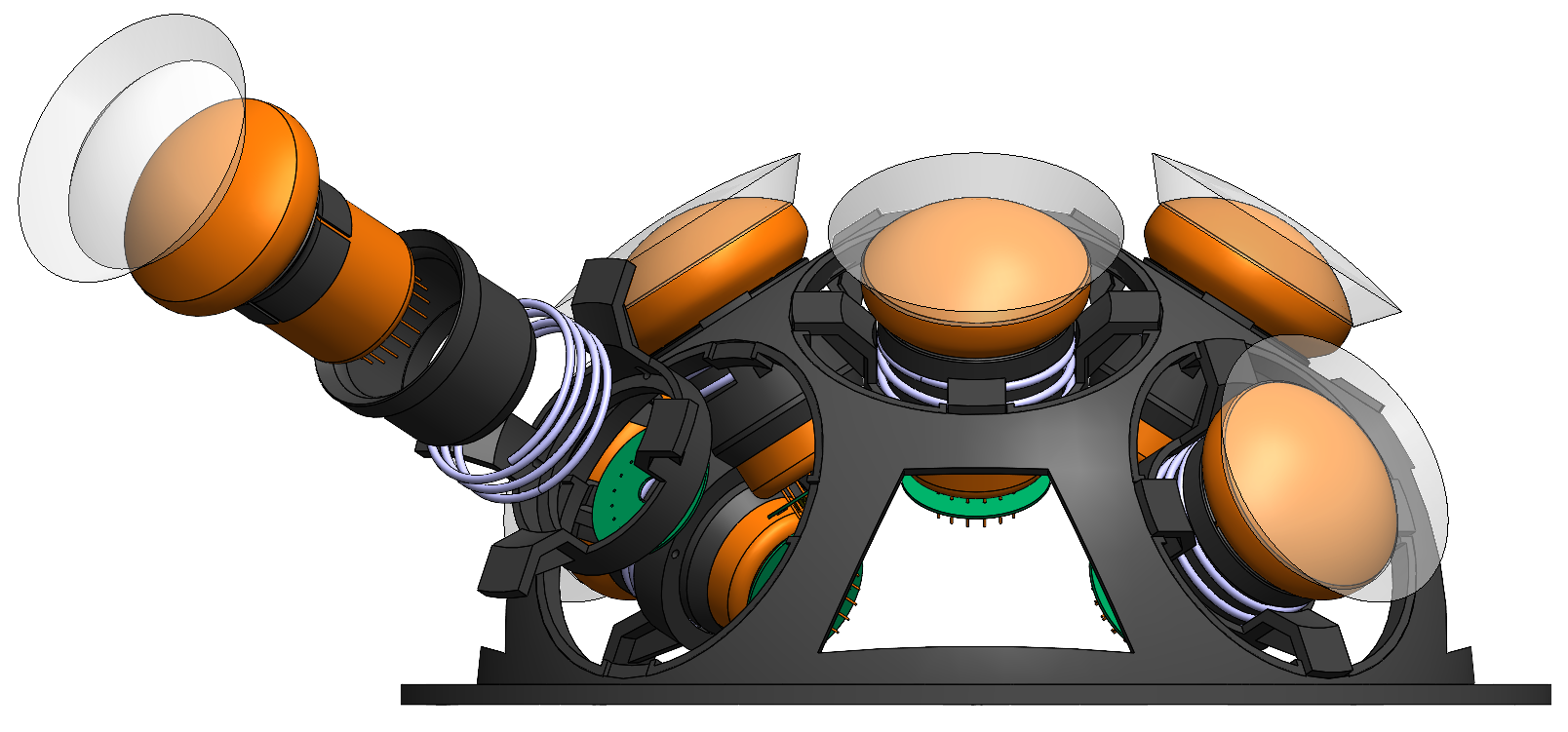} 
        \caption{ P-ONE optical module hemisphere with photomultiplier tubes (orange).}
    \end{subfigure}
    \caption{Preliminary P-ONE optical module concepts for both the full cable segment including the glass hemispheres, detection electronics and titanium frame (a) and internal detection components of a hemisphere (b). The latter shows photomultipliers used for detecting light in the deep-sea, peripheral electronics, and mounting structure.}
    \label{fig:pom}
\end{figure}

The current design of in-module cable terminations allows a streamlined detector line design, while the measurement instruments will be enclosed in 17" glass hemispheres connected to the titanium termination frame. All electrical connections are done within this frame. A layout of the P-ONE optical module (P-OM) is shown in \cref{fig:pom}. The P-OM further hosts several photomultiplier tubes (PMTs), devices sensitive to single photons with a timing resolution on nanosecond scales. These sensors will be used to detect light from interactions of cosmic particles in the deep-sea to study particle- and astrophysics. 
This is the primary science objective of P-ONE with more details described in~\cite{Agostini_2020}. 
In addition to these physics goals, the light detected by the PMTs can be used to study optical properties of the deep-sea and monitor bioluminescence activity during the operational time of P-ONE. 
Both STRAW pathfinder missions have already been able to produce first results on optical properties and bioluminescence activity~\cite{Boehmer_2019,Bailly_2021}.

In addition to the PMTs, each P-OM will host temperature sensors and accelerometers, continuously monitoring the ocean temperature and the acceleration of instruments. In the current planning, some modules within the mooring line will be instrumented with acoustic receivers, detecting the acoustic pulses from the seafloor and calibrating the position of the array moving due to the Ocean currents of the site. Using a mechanical model of the string, this positional calibration will in turn allow for indirect, but continuous monitoring of current velocity and direction. The used technology, frequency ranges and required sensitivities for both acoustic transmitters and receivers is under investigation. 

\section{Interdisciplinary science}
An initial multi-disciplinary meeting took place in October 2021. It identified a number of possible oceanographic applications that could leverage the P-ONE infrastructure. They are presented in this section, as a non-exhaustive list.

\subsection{Bioluminescence}
\begin{figure}[h!]
    \centering
    \includegraphics[width=1.\textwidth]{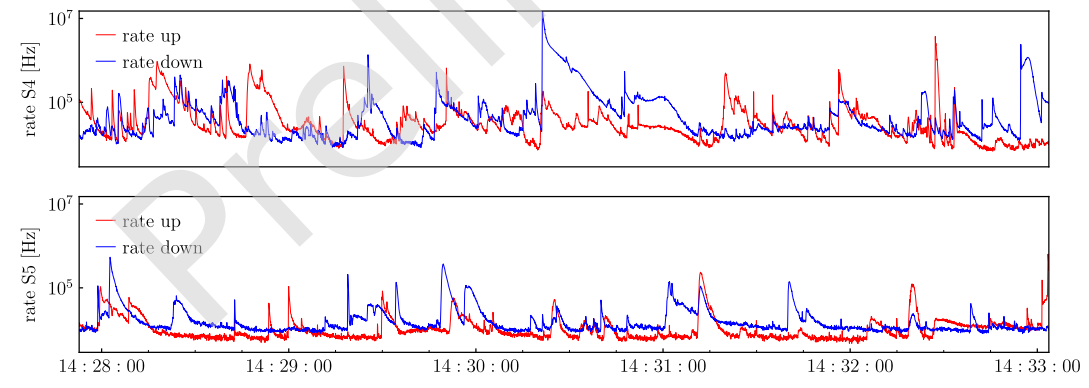}
    \caption{Preliminary STRAW photosensor rates for instruments S4/S5, each showing rates for up- and down-facing PMTs in a data period of five minutes. The visible sub-structure clearly shows varying biolumiscence activity throughout the water column sampled by the STRAW detector. The figure is a part of the full figure taken from~\cite{Bailly:2021Qi}.}
    \label{fig:biolum}
\end{figure}

With P-ONE being a highly light-sensitive detector, it will inherently allow collecting light of the deep-ocean ambient environment. 
Monitoring bioluminescence in the deep-sea has been a difficult task and the possibility of large-scale, long-term monitoring presents itself with the P-ONE experiment. 
In addition to using the photodetectors for monitoring general bioluminescence activity, the possibility to monitor the deep-sea environment using high-definition color cameras exists, 
and can be explored with further clarification of potential camera requirements for deep-ocean bioluminescence observations. 
The pathfinder experiment STRAW-b already showed this to be a promising research topic~\cite{Bailly_2021} with an exemplary plot of rates measured by its photosensor instruments shown in \cref{fig:biolum}.
With no other such observatory present in the Pacific Ocean, we anticipate P-ONE to be a strong, interdisciplinary research opportunity for various fields researching bioluminescence and related aspects of deep-sea biology.

\subsection{Acoustic tomography and monitoring}
Current planning foresees that the P-ONE array will make use of various acoustic receivers distributed within the cubic-kilometer volume of the experiment for calibration purposes.
With such a volumetric sampling of acoustic signals, the topic of acoustic tomography, that is the measurement of sound-wave propagation in water, presents itself as an exciting research opportunity.
Large-scale measurements of low-frequency, high-power acoustic beam patterns from sources in Hawaii~\cite{osti_1558213} with the P-ONE detector will allow observing the Ocean dynamics in the region between emitter and receiver. 
As such, these measurements can directly probe the temperature distribution throughout the whole water column and other climatological variations.

For acoustics, expertise and background in ocean acoustics and instrumentation would be beneficial to the project and could extend P-ONE acoustics beyond geometric calibration of the array. 
This could include acoustic monitoring of the deep-sea environment (earthquakes, mammals, wind, etc.) or making use of strong acoustic sources to investigate large-scale variation of the Ocean's speed of sound.

\subsection{Deep ocean dynamics and thermodynamics}
Since ocean currents will sway the detection mooring lines of P-ONE, one of the primary calibration targets is inferring the instrument positions at periodic time intervals. In combination with optical and acoustic signals, each instrument's position is determined with high accuracy, and will allow estimating the water current magnitude and direction. With further temperature, pressure and acceleration sensors integrated in every instrument, the three-dimensional P-ONE array will be able to track internal waves, time-dependent water current vectors, and the geothermal heat flux from the ocean floor using thermometric measurements. 

\section*{Acknowledgements}
We acknowledge the support of Ocean Networks Canada and all member institutions of the P-ONE collaboration. The document template is based on \href{https://github.com/kourgeorge/arxiv-style}{https://github.com/kourgeorge/arxiv-style}.

\printbibliography

\end{document}